\documentclass[aps,prl,twocolumn,showpacs,superscriptaddress,groupedaddress]{revtex4-1}  
\usepackage{graphicx}  
\usepackage{dcolumn}   
\usepackage{bm}        
\usepackage{amssymb}   
\usepackage{amsmath}
\usepackage{xspace}
\usepackage{epsfig}
\usepackage{times}
\usepackage{graphicx}
\usepackage{bbold}
\usepackage{float}
\usepackage[usenames,dvipsnames]{color}

\newcommand\ket[1]{| #1 \rangle}
\newcommand\bra[1]{\langle #1 |}
\newcommand\qip[2]{\langle #1 | #2 \rangle}

\begin{document}

\widetext

\title{Classical vs Quantum Games: Continuous-time Evolutionary Strategy Dynamics}
\author{Ming Lam ~Leung} \affiliation{The Chinese University of Hong Kong, Hong Kong}
\date{\today}

\begin{abstract}
This paper unifies the concepts of evolutionary games and quantum strategies. First, we state the formulation and properties of classical evolutionary strategies, with focus on the destinations of evolution in 2-player 2-strategy games. We then introduce a new formalism of quantum evolutionary dynamics, and give an example where an evolving quantum strategy gives reward if played against its classical counterpart.
\end{abstract}

\maketitle


Game theory \cite{vonN47} is a theory of decision making, which provides mathematical tools for investigating situations in which several parties make decisions according to their own interests. In a game, each player aims at maximizing their utilities or payoffs by choosing the best choice of strategies they can offer. The last few decades witnessed several new concepts of game theory, such as quantum games and evolutionary games, attracting much attention among not only game theorists, but also mathematicians, quantum physicists, economists, biologists and computer scientists.

Quantum games proposed a new perspective for the solution of game theory. In a classical game, players draw a mixed strategy from a probability distribution $p$. To modify ones strategy, each player $i$ is allowed to apply linear transformation among her own part of probability vector $p_i$. In quantum games, however, players employ quantum strategies instead of mixed strategies, over a larger set of quantum states $\rho$ in Hilbert space $\mathcal{H}$. It has been shown that quantum strategies lead to advantages over classical strategies in some particular game examples. For instances, Meyer \cite{Meyer99} quantized a coin tossing game and discovered that one may use quantum strategy against his opponent's classical strategy to win a high utility with certainty. Eisert et. al. \cite{Eisert99} proposed a new model for 2-player 2-strategy games with entanglement by quantizing prisoner's dilemma. Zhang \cite{zhang} modified their model and allowing arbitrary admissible local operations to each player, and addressed a quantitative study on the increase of equilibrium payoffs in quantum games. All previous results have been restricted to static games, without considering any form of dynamical strategy evolution.

Evolutionary game theory \cite{Mayn73,May82,We95}, innovated by the study of biological evolution, does not rely on the assumption of completely rational players but on the idea that Darwinian process of natural selection \cite{Fisher30} drives biological systems towards optimization of future reproductive success \cite{Ham94}. John Maynard Smith \cite{May82} was the first geneticist who tried to model the dynamics of biological systems in terms of game theory. He introduced a new concept called evolutionary stable strategy to explain the distribution of different phenotypes within biological populations \cite{Mayn73}. Generally, given the set of differential equations of strategy dynamics, the equilibrium states can be determined as the fixed points of the dynamical system. One particular model describing how populations evolve is called replicator dynamics, whose fixed points are found to be Nash equilibria \cite{Myer91}.

To begin with, we briefly review the notion of classical evolutionary game by the following game example. Assume two populations, Assyrian and Babylonian, are going to decide whether to trade (T) or to farm (F) for their livings. Generally trading makes good profit, but both populations need to cooperate (both choose T) in order to make successful trading, otherwise the one who performs trading earns nothing. The payoffs to players according to their strategies pair are determined by payoff matrices $A$ and $B$ (expression \ref{matrix}). Notice that the game is a symmetric game so payoff matrix $B$ is the transpose of $A$.

\begin{eqnarray}
	A =  \begin{bmatrix} 1 & 0 \\ 0.5 & 0.5 \end{bmatrix} \quad ; \quad  B = A^T = \begin{bmatrix} 1 & 0.5 \\ 0 & 0.5 \end{bmatrix}  \label{matrix}
\end{eqnarray}

Now Assyrian and Babylonian play the game repeatedly. In each stage, they have to decide how to distribute their people into a mixed proportion of trading and farming. We called their distributions of people as two mixed strategies $\vec{x}$ and $\vec{y}$, each is represented by a 2-dimensional probability vector over strategies T and F. The game is equivalent to a finite-strategy game that each player chooses strategy T or F according to probability distribution $\vec{x}$ and $\vec{y}$. The expected payoff of each player is a bilinear function of $\vec{x}$ and $\vec{y}$, given by $u^A=\vec{x}^T A \vec{y}$ and $u^B=\vec{x}^T B \vec{y}$.

A strategy pair $(\vec{x},\vec{y})$ is represented by a point in a 2-dimensional affine subspace $\mathcal{S}$ in $[0,1]^4$. Assuming that the time step between two stage games is infinitesimally small and the game is played infinitely many times, the mixed strategies $\vec{x}(t)$ and $\vec{y}(t)$ can be treated as continuous vector functions of time $t$. The product strategy is a time-dependent vector $\vec{p}(t)=\vec{x}(t) \times \vec{y}(t)$ in the strategy space $\mathcal{S}$, which trajectory draws an evolution path of their mixed strategy distribution.

In this Trading-Farming game, here are 3 dynamical equilibria: both players choose T, both players choose F, or both choose T with probability $1/2$. We called the first two states as pure equilibria (PE), and the third one as internal equilibrium (IE), which is an equilibrium probability vector with no zero entry. Since these 3 strategy pairs lead both players reaching local optimal payoff, they would not have the intention to modify their strategies once they reached one of the equilibrium states. On the contrary, if they are not using equilibrium strategies in current stage, there exists an alternative strategy for each player to earn more. As to maximize future payoffs, a strategy evolves with time subject to payoff difference. Therefore, the strategy evolution can be captured by certain kind of evolutionary dynamics.

One of the most commonly mentioned evolutionary dynamics is called \emph{replicator dynamics}, which was well-established by Bomze (1983) and applied to explain animal evolution behaviors \cite{Bom83}. Formally, following replicator dynamics, the evolution of Assyrian's and Babylonian's strategy is determined by the following differential equations:
\begin{eqnarray}
\nonumber {dx_i \over dt} = \gamma x_i ( (A\vec{y})_i - \vec{x}^T A \vec{y}) \\
\nonumber {dy_i \over dt} = \gamma y_i ( (\vec{x}^T B)_i - \vec{x}^T B \vec{y}) 
\end{eqnarray}
Intuitively, if a pure strategy (say T) yields a better payoff compared to the current mixed strategy $\vec{x}$, Assyrian will gradually expand the probability share of using T in proportion to the payoff difference, where $\gamma$ is the rate of evolution.

Below are some general properties of classical 2-player 2-strategy evolutionary games. Generally, with payoff matrices $A =  \begin{bmatrix} a & b \\ c & d \end{bmatrix}$ and $B = \begin{bmatrix} a' & b' \\ c' & d' \end{bmatrix}$, let the players employ mixed strategies $\vec{x} = (x, 1-x)$ and $\vec{y} = (y, 1-y)$, so Assyrian will choose T with probability $x$ and Babylonian will choose T with probability $y$. For every strategy pair $(x,y)$, we let $\vec{v}(x, y)$ be a velocity vector function subject to replicator dynamics. For any 2-player 2-strategy game, there is at most 1 internal equilibrium. An internal equilibrium is uniquely determined to be $(x^*, y^*) = ({d'-c' \over a'-b'-c'+d'},{d-b \over a-b-c+d})$ if it exists. Based on its position, the probabilistic strategy space can be divided into 4 quadrants, with the internal equilibrium $(x^*, y^*)$ locates at centre. We claim that for any two internal points within the same quadrant, the sign of each component of their velocity vectors should be the same. For any two internal points and located in different quadrants, their velocity vectors have at least one component different in sign. These properties can be shown by the bilinearity of the expected payoff function, where the payoff has to be monotonic increasing or decreasing if we fix the strategy of one player. Thus it is obvious that for any 2-player 2-strategy game without internal equilibrium, an evolving strategy will always collapse to a state on boundary, which means at least one player ends up with a dominant pure strategy by the end of evolution.
\begin{figure}
\includegraphics[scale=0.4]{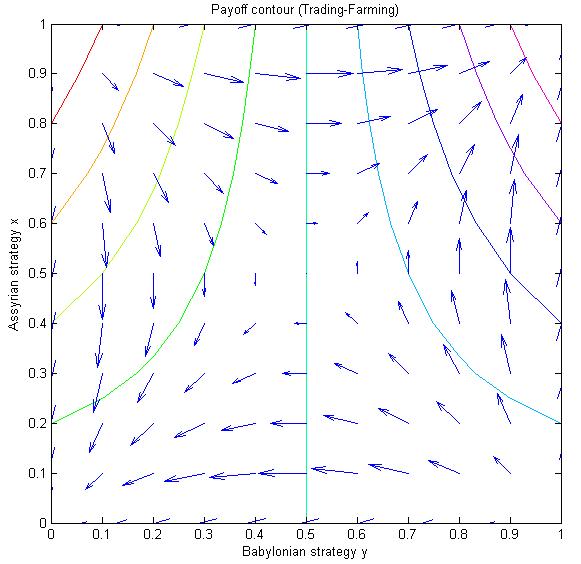}
\caption{\label{fig1} A contour plot of strategy space of a Trading-Farming game.}
\vspace{-0.1in}
\end{figure}

Next, we put forth our discussions to symmetric games. For any 2-player 2-strategy symmetric game, the payoff matrices can be expressed in $4$ parameters as below:
\begin{eqnarray}
\label{gmatrix}
	A =  \begin{bmatrix} a & b \\ c & d \end{bmatrix} \qquad ; \qquad  B = A^T = \begin{bmatrix} a & c \\ b & d \end{bmatrix}  
\end{eqnarray}
In particular, our previous Trading-Farming game has parameters $a > c = d > b$, and the famous Prisoner-Dilemma game is another example with $b > d > c > a$. A game with $a > c$ and $b > d$ (or $a < c$ and $b < d$) has no mixed equilibrium and only one pure equilibrium, then the strategy evolution will always converge to that pure equilibrium. For symmetric games with single internal equilibrium, we classify them into two categories. One category (type I) includes symmetric games like Hawk-Dove, which have parameters $a < c$ and $b > d$. Another category (type II) includes symmetric games like Trading-Farming, which have parameters $a > c$ and $b < d$.

In a type I symmetric game, assuming both players follow same evolutionary dynamics, if they start at same initial mixed strategy, their strategies evolve towards the internal equilibrium. Otherwise, the player with larger initial share of strategy T ultimately converges to pure strategy T and the other player converges to pure strategy F. One can deduce the above properties by the fact that an evolving strategy never moves across the diagonal $x=y$ in the $2$-dimensional strategy space, due to the symmetric property of evolutionary dynamics, i.e. $v_x(x,y) = v_y(y,x)$. Therefore, any unequal strategy pair will finally converge to either TT or FF through evolution.

\begin{figure}
\includegraphics[scale=0.4]{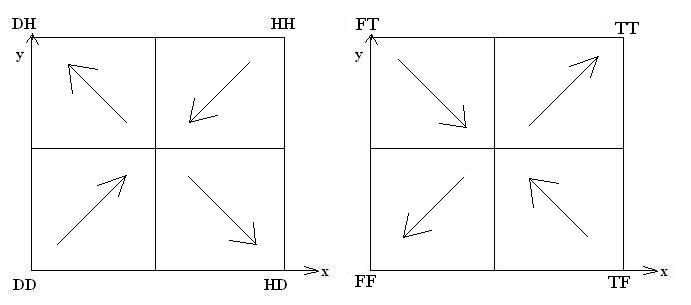}
\caption{\label{fig2} Type I and Type II symmetric game.}
\vspace{-0.1in}
\end{figure}

The result for type II symmetric games is completely different. Let H and D be the choices of strategies, given an internal equilibrium $(x^*, y^*)$, for any point in the region $x > x^*$ and $y < y^*$, the evolution converges to HD. Within the region $x < x^*$ and $y > y^*$, the evolution converges to DH. For the other two regions, it can be shown that if a strategy at $(x,y)$ converges to HD, a point $(x-\alpha, y+\beta)$ also converges to HD, for all $0 < \alpha < x$ , $0 < \beta < 1-y$. Similarly, if a strategy at $(x,y)$ converges to DH, a point $(x + \alpha, y - \beta)$ also converges to DH for all $0 < \alpha < 1-x$ , $0 < \beta < y$. Combining the results above, given any evolutionary symmetric game $G$ under replicator dynamics, the strategy space can be divided into $2$ open regions, where each pure equilibrium lies in each regions, and the internal equilibrium lies on the boundary. Then any initial strategy inside a region converges to pure equilibrium contained in that region, and any strategy located on boundaries between regions converges to the internal equilibrium.

Notice that replicator dynamics is not the only possible dynamical system describing continuous strategy evolution. As soon as the evolutionary dynamics satisfies adjustment property \cite{Sw93} : ${d \over dt} \vec{x}^T A \vec{y} \ge 0$, the arguments in this letter follows.

Now we consider quantum evolutionary games which allow players to use strategies quantum mechanically. In a classical game, each player has a strategy set $\mathcal{S}$, and the mixed strategy distribution is represented by probability vector $\vec{p}=\vec{x} \times \vec{y}$. The expected payoffs are given by $u^A(\vec{p}) = \sum_{ij} x_i y_j A_{ij}$ and $u^B(\vec{p}) = \sum_{ij} x_i y_j B_{ij}$. In a quantum game, each player $i$ now has a Hilbert space $\mathcal{H}_i = span\{s_i: s_i\in \mathcal{S}_i\}$. A quantum strategy can be represented by a quantum state $\ket{\psi}=\sum_{ij} \sqrt{x_i y_j}\ket{ij}$. Here a quantum state is a generalization of classical probability distribution. Given a quantum state $\ket{\psi}$, players receive their payoffs by performing a quantum measurement on computational basis $\mathcal{S}$, where the payoff function of each player is defined to be $u^A(\ket{\psi}) = \sum_{ij} |\qip{ij}{\psi}|^2 A_{ij}$ and $u^B(\ket{\psi}) = \sum_{ij} |\qip{ij}{\psi}|^2 B_{ij}$. In evolutionary games, a classical player can shuffle the probability distribution over her strategy space, a quantum player can perform operation on quantum state $\ket{\psi}$ by applying admissible super-operators in her local space $\mathcal{H}_i$.

Since classical and quantum strategies locate in different strategy space, their evolutionary behaviours would be possibly different. Zhang \cite{zhang} shows that a quantum equilibrium strategy $\ket{\psi}$ always induces a classical equilibrium distribution through measurement. More precisely, if $\ket{\psi}$ is a quantum Nash equilibrium, then $\vec{p}$ defined by $p(s) = |\qip{s}{\psi}|^2$ is essentially a classical Nash equilibrium. Since Nash equilibrium is a sufficient condition for a strategy to remain stationary under replicator dynamics \cite{Fri91}, the above theorem implies that a quantum equilibrium state $\ket{\psi^*}$ always implies a classical evolutionary fixed point at corresponding position $\vec{p}^*$ given by $p(s) = |\qip{s}{\psi}|^2$.

Now we use quantum replicator dynamics as an example to illustrate the evolution of non-equilibrium strategy. For classical replicator dynamics, a player tends to increase the probability share $x_i$ of a particular strategy $i$ only when the payoff of pure strategy $s_i$ is higher than the payoff of mixed strategy $\vec{x}$. The rate of change of $x_i$ is proportional to the net increase of payoff and its share following the expression below:
\begin{eqnarray}
\nonumber {d\vec{x}_i \over dt} &= \sum_{jk} x_j y_k (A_{ik}-A_{jk})  \\
                                                      &= x_i (u^A(s_i,\vec{y}) - u^A(\vec{x},\vec{y}))
\end{eqnarray}
Assume that Babylonian uses mixed strategy $\vec{y}$, here we denote the expected payoff of Assyrian using pure strategy $s_i$ to be $u^A(s_i,\vec{y})=\sum_j A_{ik} y_k$ , and its expected payoff using mixed strategy $\vec{x}$ to be $u^A(\vec{x},\vec{y})=\sum_{jk} A_{jk} x_j y_k$. We name the function $v^{(x)}_i=u^A(s_i,y)-u^A(x,y)$ be the partial velocity of evolutionary strategy $s_i$. Then we denote the diagonal matrix $V^{(x)}$ with $V^{(x)}(i,i) = v^{(x)}_i$ to be the velocity operator. Therefore, ${d \over dt} \vec{x}(t) = V^{(x)}\vec{x}(t)$.

\begin{figure}
\includegraphics[scale=0.3]{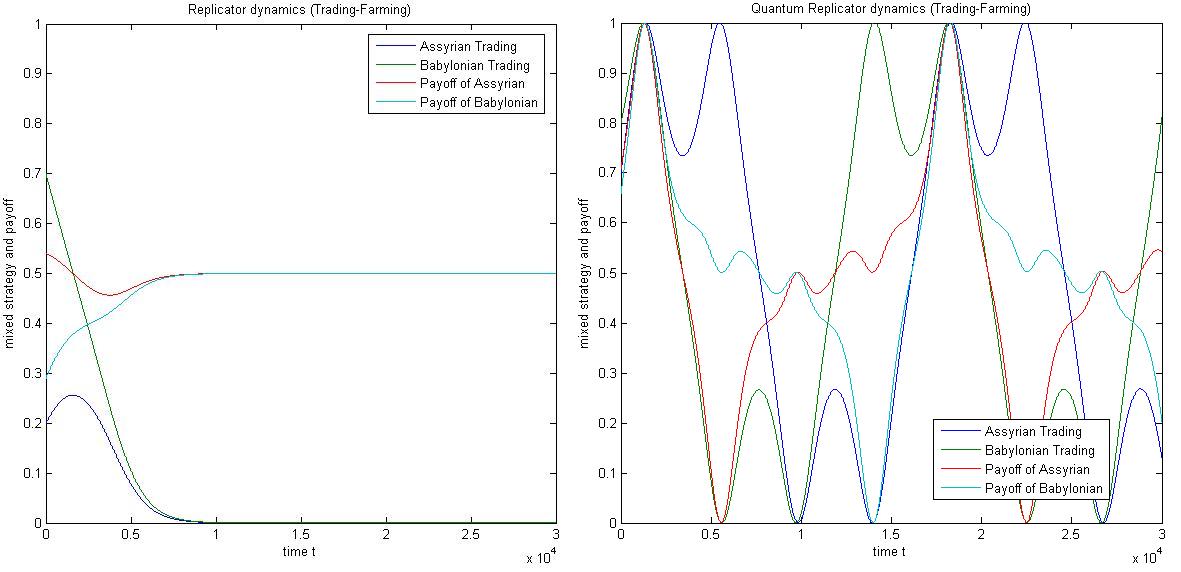}
\caption{\label{fig3} Classical (left) and quantum (right) replicator dynamical evolution in Trading-Farming game.}
\vspace{-0.1in}
\end{figure}

Consider the evolution of joint strategy $\vec{p}=\vec{x} \times \vec{y}$, since ${d \over dt} \vec{p} = \vec{x} \times {d \over dt} \vec{y} + {d \over dt} \vec{x} \times \vec{y} = (I \otimes V^{(y)} + V^{(x)} \otimes I)\vec{p}= V \vec{p}$, where $V=I \otimes V^{(y)} + V^{(x)} \otimes I$ is called the joint velocity operator which is also a diagonal matrix.

With analog to classical regime, the evolution of quantum strategy also depends on payoff differences between the target and current strategies. In this case, the quantum partial velocity $v^{(x)}$ is given by $v^{(x)}_i =  u^A(\ket{i}\ket{\psi_y})- u^A(\ket{\psi})$, where $\ket{\psi_y}=Tr_x(\ket{\psi})$. In quantum mechanics, the evolution of a quantum state is generated by a \emph{Hamiltonian} $H$, and the rate of change of a state is described by Schr\"{o}dinger equation:\begin{eqnarray}
i\hbar {d \over dt}\ket{\psi} = H \ket{\psi}
\end{eqnarray}

Here the quantum generating function of evolution is given by $-i H/\hbar$. Following replicator dynamics, given a quantum strategy $\ket{\psi}$, the Hamiltonian generator $H^{(x)}$ of Assyrian following \emph{quantum replicator dynamics} is defined as below.

\begin{eqnarray*}
H^{(x)}(\ket{\psi}) = \sum_{i,j,k} - {i \over \bar{h}} \Bigl( u^A(\ket{i}\ket{\psi_y}) - u^A(\ket{j}\ket{\psi_y}) \Bigr) \qip{k}{\psi_y} \ket{i}\bra{j}
\end{eqnarray*}
Then the product Hamiltonian $H=I \otimes H^{(y)} + H^{(x)} \otimes I$ drives quantum strategies of players to evolve, which itself is state-dependent and thus time-dependent during evolution. For any later time $t>0$, the quantum strategy $\rho(t)$ evolves based on the initial state $\ket{\psi(0)}$ and Hamiltonian generator $H(\ket{\psi})$ according to the following:
\begin{eqnarray}
\ket{\psi(t)}=U(0,t)\ket{\psi(t)} =e^{-\int_0^t i H(\ket{\psi(t')})dt'}\ket{\psi(0)}
\end{eqnarray}

Quantum replicator dynamics is only applicable on quantum strategy but not classical strategy. Different from its classical counterpart, a quantum strategy may evolve drastically even if its classical counterpart is a dominant pure equilibrium.
%
For example, consider a 2-player symmetric game $G$ with payoff matrices $A$ and $B$ define as expression \ref{gmatrix}. In classical regime, let $\vec{x}(t)=(\cos^2\theta \quad \sin^2 \theta)^T$ and $\vec{y}(t)=(\cos^2\phi \quad \sin^2 \phi)^T$ be the mixed strategy of Assyrian and Babylonian use at time $t$. For initial strategies $\vec{x}(0)=(\cos^2\theta_0 \quad \sin^2 \theta_0)^T$ and $\vec{y}(0)=(\cos^2\phi_0 \quad \sin^2 \phi_0)^T$. Under classical replicator dynamics, a strategy evolve with time as below.
\begin{align}
\vec{x}(t)&=\vec{x}(0)+ \int_0^t  \gamma\delta(\phi) \sin^2\theta \cos^2\theta \begin{bmatrix} 1\\ -1 \end{bmatrix}  dt'  \label{s17} 
\end{align}
where $\delta(\theta) = (a-c) \cos^2\theta + (b-d)\sin^2 \theta$ and $\delta(\phi) = (a-c) \cos^2\phi + (b-d)\sin^2 \phi$. The above expressions are deduced by substituting ${d\theta \over dt} = -\delta(\phi)\sin2\theta/4$ and ${d\phi \over dt} = -\delta(\theta)\sin2\phi/4$ into the replicator dynamics expression.

On the other hand, players may employ quantum strategies following quantum replicator dynamics. Let $\ket{\psi} = (e^{i\alpha} \cos\theta \quad e^{-i\alpha} \sin\theta)^T$ and $\ket{\zeta} = (e^{i\alpha} \cos\phi \quad e^{-i\alpha} \sin\phi)^T$ be the quantum strategies of Assyrian and Babylonian respectively. Their strategy evolutions are governed by the Hamiltonian generators $H_A$ and $H_B$, where
\vspace{-0.1in}
\begin{eqnarray*}
H_A(\ket{\psi}) = \sum_{a,b} \gamma \qip{a}{\psi} \Bigl( u^A(\ket{a}\ket{\zeta}) - u^A(\ket{b}\ket{\zeta}) \Bigr) \ket{a}\bra{b} 
\end{eqnarray*}
\vspace{-0.1in}
Therefore,
\begin{align}
\nonumber \ket{\psi(t)}&=e^{-\int_0^t iH_Adt'}\ket{\psi(t)} \\ 
                    &=\ket{\psi(0)}+ \int_0^t \gamma \delta(\phi) \begin{bmatrix} -e^{-i\alpha} \sin\theta \\ e^{i\alpha} \cos\theta\end{bmatrix} dt'  \label{s19} 
\end{align}

Comparing equations \ref{s17} and \ref{s19}, we see that quantum replicator dynamical evolution is different from its classical counterpart. One significant difference is that, the classical strategy will converge whenever $\theta$ or $\phi \rightarrow 0$ (which leads to a classical pure strategy), since $\sin^2 0=0$ leads to a static evolution; but the quantum strategy keeps evolving in the complex space even if $\theta$ or $\phi \rightarrow 0$.

Particularly, in Prisoner-Dilemma game, a pair of classical evolving strategies always converge to pure strategy Defect, but a pair of evolving quantum strategies keep oscillating within a cycle and never converge (See figure 4). Such a difference is due to the Hilbert space of quantum strategy form a spherical manifold instead of an affine 2D-plane with borders. When a quantum strategy vector reaches pure strategy Defect, it keeps moving towards the negative quadrant and never stops at s. This makes the quantum state evolving in a circular loop and never converges to any equilibrium.

\begin{figure}
\includegraphics[scale=0.3]{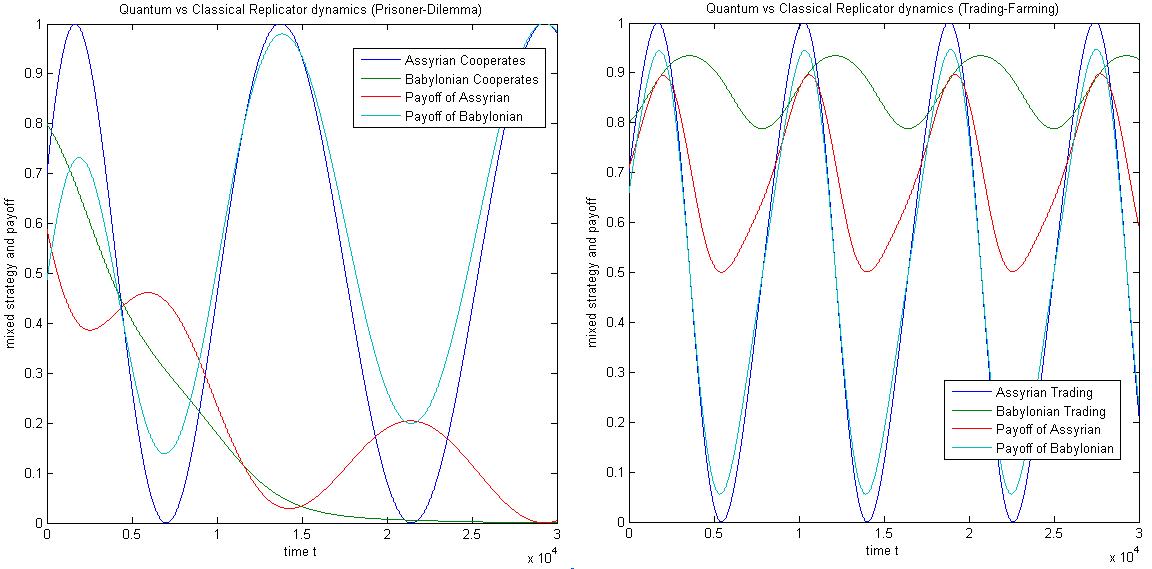}
\caption{\label{fig4} Classical vs Quantum replicator dynamical evolution in Prisoner-Dilemma (left) and Trading-Farming (right). Assyrian employs quantum evolving strategy, Babylonian employs classical evolving strategy.}
\vspace{-0.1in}
\end{figure}

Previously we consider both players playing quantum evolving strategies. One may be curious if quantum dynamics provides us some advantages in a unfair game against its classical counterpart. In iterated Prisoner-Dilemma based on replicator dynamics, quantum does not provide us any benefit. On the other hand, in Trading-Farming game, regardless of its less advantageous initial position, employing a quantum replicator dynamics ensures quantum player (Assyrian) to achieve a better payoffs compared to classical replicator dynamics (Refer to figure \ref{fig3} and \ref{fig4}).

The author would like to acknowledge his supervisor, Professor Shengyu Zhang, for fruitful discussions and offering continuing advice and giving encouraging guidance.


\end{document}